\documentclass{emulateapj}
\usepackage{apjfonts}

\shorttitle{SHARC-2 $350\,\mu$m OBSERVATIONS OF DISTANT SMGs}
\shortauthors{KOV\'{A}CS ET AL.}

\begin{document}

\title {SHARC-2 350$\,\mu$\lowercase{m} Observations of Distant Submillimeter Selected Galaxies}

\author{A. Kov\'{a}cs\altaffilmark{1}, S. C. Chapman\altaffilmark{1}, C. D. Dowell\altaffilmark{1}, A. W. Blain\altaffilmark{1}, R. J. Ivison\altaffilmark{2}, I. Smail\altaffilmark{3}, and T.G. Phillips\altaffilmark{1}}

\altaffiltext{1}{California Institute of Technology, Mail Code 320-47, 1200 E California Blvd, Pasadena, CA 91125}
\altaffiltext{2}{Royal Observatory, Blackford Hill, Edinburgh EH9 3HJ, UK}
\altaffiltext{3}{Institute for Computational Cosmology, Durham University, South Road, Durham DH1 3LE, UK}

\journalinfo{The Astrophysical Journal, 650:592-603, 2006 {\rm October} 20}
\submitted{Received 2006 January 26; Accepted 2006 April 27}

\begin{abstract}
We present $350\,\mu$m observations of 15 Chapman et al.\ submillimeter galaxies (SMGs) with radio counterparts and optical redshifts. We detect 12 and obtain sensitive upper limits for three, providing direct, precise measurements of their far-infrared luminosities and characteristic dust temperatures. With these, we verify the linear radio--far-infrared correlation at redshifts of $z\sim 1$--3 and luminosities of $10^{11}$--$10^{13}\,L_{\odot}$, with a power-law index of $1.02\pm0.12$ and rms scatter of $0.12$\,dex. However, either the correlation constant $q$ or the dust emissivity index $\beta$ is lower than measured locally. The best fitting $q\simeq2.14$ is consistent with SMGs being predominantly starbust galaxies, without significant AGN contribution, at far-infrared wavelengths. Gas-to-dust mass ratios are estimated at $54^{+14}_{- 11}\:\bigl(\kappa_{850\,\mu{\rm m}}/0.15$\,m$^2$\,kg$^{-1}\bigr)$, depending on the absoption efficiency $\kappa_\nu$, with intrinsic dispersion $\simeq 40\%$ around the mean value. Dust temperatures consistent with $34.6\pm3\,$K$\:(\beta/1.5)^{-0.71}$, at $z\sim1.5$--3.5, suggest that far-infrared photometric redshifts may be viable, and perhaps accurate to $10\%\lesssim dz/(1+z)$, for up to $80\%$ of the SMG population in this range, if the above temperature characterizes the full range of SMGs. However, observed temperature evolution of $T_d\propto (1+z)$ is also plausible and could result from selection effects. From the observed luminosity-temperature ($L$-$T$) relation, $L\propto T_{\rm obs}^{2.82\pm0.29}$, we derive scaling relations for dust mass versus dust temperature, and we identify expressions to inter-relate the observed quantities. These suggest that measurements at a single wavelength, in the far-infrared, submillimeter or radio wave bands, might constrain dust temperatures and far-infrared luminosities for most SMGs with redshifts at $z\sim 0.5$--4.
\end{abstract}


\maketitle

\section{Introduction}

Ever since the first submillimeter-selected galaxy (SMG) samples debuted from SCUBA $850\,\mu$m surveys \citep{Smail1997, Barger1998, Hughes1998, Eales1999}, the nature of the SMG population has been the focus of attention for the galaxy formation community, because the $850\,\mu$m selection is expected to pick similar sources almost independently of redshift ($z\sim 1$--8), due to a negative $K$-correction that essentially compensates for the loss of flux from increasing distance. This allows unbiased, luminosity-selected studies of galaxy formation. The hunger for information on these sources spurred a flurry of follow-up studies at all wavelengths, long and short. Since then many of these sources have been identified at optical, UV \citep{Borys2003, Chapman2003, Webb2003} and radio wavelengths \citep{Smail2000, Ivison2002}, providing accurate positions, which allowed optical redshift measurements \citep{Chapman2003, Chapman2005}. As a result we now know that these massive galaxies, with redshifts distributed around $z \simeq 2.3$, are enshrouded with such quantities of dust that they often lie hidden at optical wavelengths, and therefore constitute a distinct population from the galaxies selected by optical surveys. 

More recently, longer wavelength submillimeter surveys, at 1100 and at 1200\,$\mu$m \citep{Laurent2005,Greve2004}, added to the pool of available information. However, the close proximity of the SCUBA, Bolocam, and MAMBO wavelengths on the Rayleigh-Jeans side of the spectral energy distribution (SED) does not allow for an effective constraint on the thermal far-infrared spectral SEDs at the relevant redshifts. Nor do the latest results from the {\it Spitzer Space Telescope} provide powerful constraints, since at the shorter mid-infrared wavelengths the emission is dominated by polycyclic aromatic hydrocarbons (PAHs) and a minority population of hot dust. For these reasons, the best estimates of the characteristic temperatures and the integrated luminosities, to date, have relied on the assumption that the local radio to far-infrared correlation \citep{Helou1985, Condon1992, Yun2001} can be extended into the distant universe. There are hints that this may be appropriate \citep{Garrett2002, Appleton2004}, but the assumption has remained largely unchecked.

Shorter wavelength submillimeter measurements, sampling near the rest-frame peak of the emission, are thus essential to provide firm constraints to the far-infrared SED. Here we present results at $350\,\mu$m, using the second-generation Submillimeter High Angular Resolution Camera (SHARC-2; \citealt{Dowell2003}) at the Caltech Submillimeter Observatory (CSO). From these we derive the first direct measures of dust temperatures and far-infrared luminosities for a sample of SMGs, testing the radio to far-infared correlation. We also attempt to constrain dust emission properties and investigate the implications of our findings for the viability of photometric redshifts based on far-infrared and radio measurements. Finally, we present a range of useful scaling relations that may apply to the SMG population.

\section{SHARC-2 350$\,\mu$\lowercase{m} Observations}

\begin{figure*}
\includegraphics[width=\textwidth]{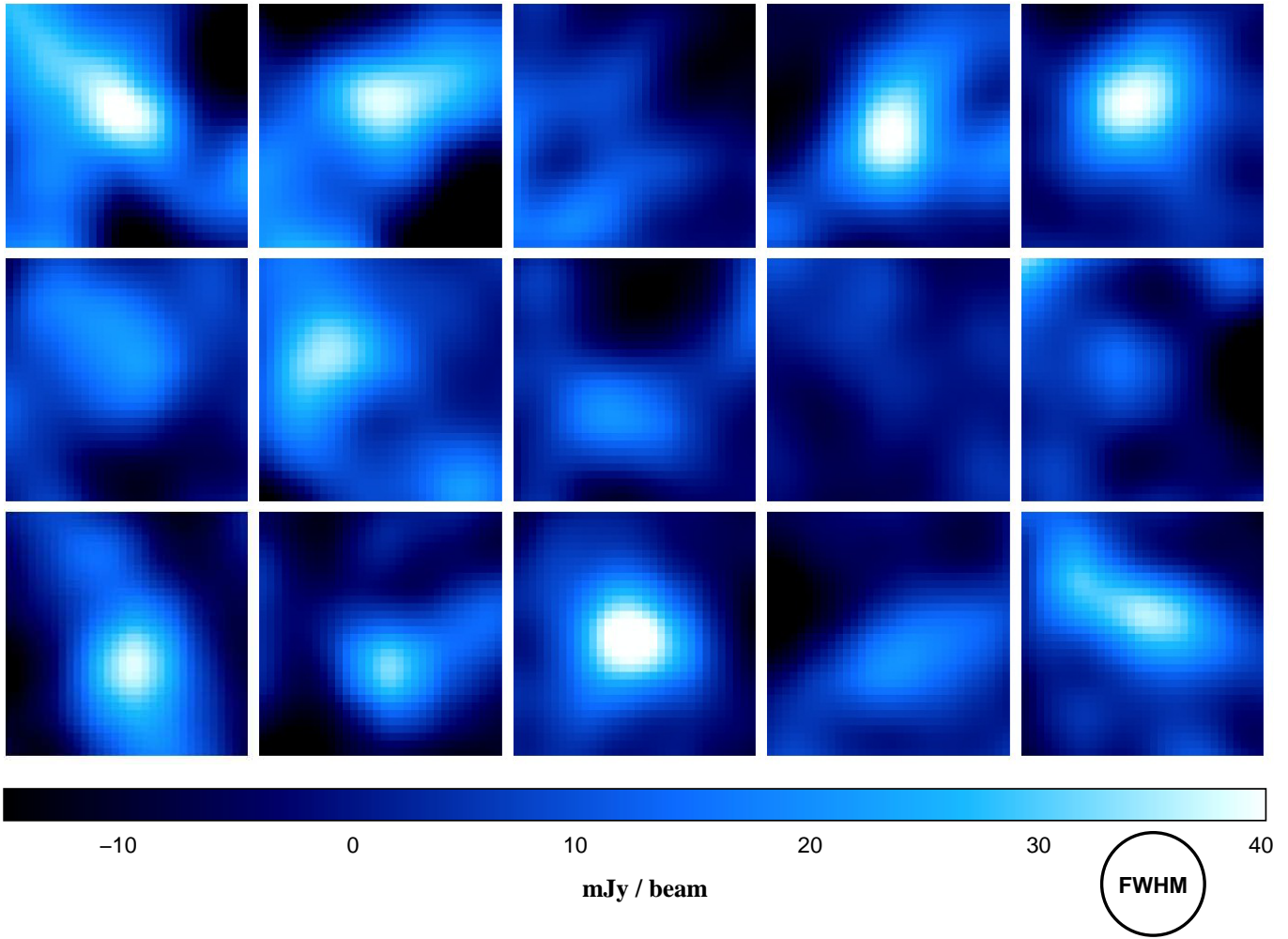}
\caption{
30''$\times$30'' SHARC-2 $350\,\mu$m thumbnail images of the SMGs observed (sources 1--15 in Tables~\ref{tab:fluxes}-\ref{tab:best-derivatives}) shown in reading order ({\it left to right, top to bottom}). Images are centered on the radio positions, and displayed on the same flux scale for comparison. The $12"\!\!\!.4$ FWHM effective beams of the optimally filtered (convolved) images are also indicated. Sources 3, 9, and 10 are not detected. 
} 
\label{fig:maps}
\end{figure*}

We conducted follow-up observations of SCUBA $850\,\mu$m detected sources with radio identifications and optical redshifts \citep{Chapman2003, Chapman2005}. Seven of the 15 targets were hand-picked on the basis of their predicted bright $350\,\mu$m fluxes ($S_{350\,\mu{\rm m}} > 45$\,mJy from \citealt{Chapman2005}), while the remaining were selected at random to partially compensate for any selection bias.

\begin{deluxetable*}{llr@{,}lcr@{~~$\pm$}lr@{~~$\pm$}lr@{~~$\pm$}lr@{~~$\pm$}lr@{~~$\pm$}l}
\tablewidth{\textwidth}
\tablecaption{Summary of Observations\label{tab:fluxes}}
\tablecolumns{15}

\tablehead{ 
  \colhead{} & \colhead{} & \multicolumn{2}{c}{Offset\tablenotemark{a}} & \colhead{}
  & \multicolumn{2}{c}{} & \multicolumn{2}{c}{} & \multicolumn{2}{c}{} & \multicolumn{2}{c}{} & \multicolumn{2}{c}{} \\
\colhead{} & \colhead{} & \multicolumn{2}{c}{(J2000.0)} & \colhead{}
  & \multicolumn{2}{c}{$S(350\,\mu$m)} & \multicolumn{2}{c}{$S(850\,\mu$m)} & \multicolumn{2}{c}{$S(1100\,\mu$m)} & \multicolumn{2}{c}{$S(1200\,\mu$m)} & \multicolumn{2}{c}{$S(1.4\,$GHz)} \\
  \colhead{ID} & \colhead{Name} & \multicolumn{2}{c}{(arcsec)} & \colhead{$z$}
  & \multicolumn{2}{c}{(mJy)} & \multicolumn{2}{c}{(mJy)} & \multicolumn{2}{c}{(mJy)} & \multicolumn{2}{c}{(mJy)} & \multicolumn{2}{c}{($\mu$Jy)} 
}

\startdata
1  & SMM J030227.73+000653.5       & +1.0&+0.8 & 1.408 & 42.2&9.8     &  4.4 & 1.3 &    \multicolumn{2}{c}{\nodata}      &    \multicolumn{2}{c}{\nodata}    &   217 & 9    \\
2  & SMM J105207.49+571904.0       & $- 1.0$&+3.7 & 2.689 & 38.0&7.2     &  6.2 & 1.6 &    \multicolumn{2}{c}{\nodata}      &(0.4 & 0.8)& 277.8 & 11.9 \\
3 \tablenotemark{b}
   & SMM J105227.77+572218.2       &  \multicolumn{2}{c}{\nodata}  &1.956 & (11.3&6.7)   &  7.0 & 2.1 &   \phn 5.1&1.3 \tablenotemark{b}
                                                                                                               & 3.1 & 0.7 &  40.4 & 9.4 \tablenotemark{c} \\
4 \tablenotemark{b}
   & SMM J105230.73+572209.5       & +3.3&$- 1.8$ & 2.611 & 41.0&6.8     & 11.0 & 2.6 &   \phn 5.1&1.3 \tablenotemark{b}   
                                                                                                               & 2.9 & 0.7 &  86.3 & 15.4 \tablenotemark{c} \\
5  & SMM J105238.30+572435.8       & +1.4&+2.5 & 3.036 & 40.5&6.5     & 10.9 & 2.4 & 4.8 & 1.3   & 4.8 & 0.6 &  61.0 & 22.0 \tablenotemark{c} \\
6  & SMM J123600.15+621047.2	   & $- 1.4$&+2.0 & 1.994 & 22.3&6.3     &  7.9 & 2.4 &    \multicolumn{2}{c}{\nodata}      &    \multicolumn{2}{c}{\nodata}    &   131 & 10.6 \\
7  & SMM J123606.85+621021.4       & +6.7&+3.5 & 2.509 & 35.1&6.9     & 11.6 & 3.5 &    \multicolumn{2}{c}{\nodata}      &    \multicolumn{2}{c}{\nodata}    &  74.4 & 4.1  \\
8  & SMM J131201.17+424208.1       &  \multicolumn{2}{c}{\nodata}  & 3.405 & 21.1&7.7     &  6.2 & 1.2 &    \multicolumn{2}{c}{\nodata}      &    \multicolumn{2}{c}{\nodata}    &  49.1 & 6.0  \\
9  & SMM J131212.69+424422.5       & +2.2&$- 4.2$ & 2.805 & (3.7&4.4)    &  5.6 & 1.9 &	  \multicolumn{2}{c}{\nodata}      &    \multicolumn{2}{c}{\nodata}    & 102.6 & 7.4  \\
10 & SMM J131225.73+423941.4       &  \multicolumn{2}{c}{\nodata}  & 1.554 &(14.7&7.4)   &  4.1 & 1.3 &	  \multicolumn{2}{c}{\nodata}      &    \multicolumn{2}{c}{\nodata}    & 752.5 & 4.2  \\
11 & SMM J163631.47+405546.9       & $- 1.3$&$- 4.0$ & 2.283 & 38.3&5.5     &  6.3 & 1.9 &	  \multicolumn{2}{c}{\nodata}      &(1.1 & 0.7)&    99 & 23   \\
12 & SMM J163650.43+405737.5       & $- 1.0$&$- 1.7$ & 2.378 & 33.0&5.6     &  8.2 & 1.7 &	  \multicolumn{2}{c}{\nodata}      & 3.1 & 0.7 &   221 & 16   \\
13 & SMM J163658.19+410523.8       & +1.1&$- 1.2$ & 2.454 & 45.2&5.3     & 10.7 & 2.0 &	  \multicolumn{2}{c}{\nodata}      & 3.4 & 1.1 &    92 & 16   \\
14 & SMM J163704.34+410530.3       & $- 1.8$&$- 4.4$ & 0.840 & 21.0&4.7     & 11.2 & 1.6 &	  \multicolumn{2}{c}{\nodata}      &(0.8 & 1.1)&    45 & 16   \\
15 & SMM J163706.51+405313.8       & $- 2.3$&+2.4 & 2.374 & 36.1&7.7     & 11.2 & 2.9 &	  \multicolumn{2}{c}{\nodata}      & 4.2 & 1.1 &    74 & 23   \\
(16) & SMM J163639.01+405635.9     &  \multicolumn{2}{c}{\nodata}  & 1.495 &     \multicolumn{2}{c}{\nodata}        &  5.1 & 1.4 &    \multicolumn{2}{c}{\nodata}      & 3.4 & 0.7 &   159 & 27   \\
(17)\tablenotemark{d} 
   & SMM J105201.25+572445.7	   &  \multicolumn{2}{c}{\nodata}  & 2.148 & 24.1&5.5     &  9.9 & 2.2 & 4.4 & 1.3   & 3.4 & 0.6 &  72.1 & 10.2 \\
(18) & SMM J105158.02+571800.2     &  \multicolumn{2}{c}{\nodata}  & 2.239 &     \multicolumn{2}{c}{\nodata}        &  7.7 & 1.7 &    \multicolumn{2}{c}{\nodata}      & 2.9 & 0.7 &  98.1 & 11.6 \\
(19)\tablenotemark{d,e} 
   & SMM J105200.22+572420.2       &  \multicolumn{2}{c}{\nodata}  & 0.689 & 15.5&5.5     &  5.1 & 1.3 & 4.0 & 1.3   & 2.4 & 0.6 &  57.4 & 13.2 \\
(20)\tablenotemark{d} 
   & SMM J105227.58+572512.4       &  \multicolumn{2}{c}{\nodata}  & 2.142 & 44.0&16.0    &  4.5 & 1.3 & 4.1 & 1.3   & 2.8 & 0.5 &  39.2 & 11.4 \tablenotemark{c} \\
(21)\tablenotemark{f} 
   & SMM J105155.47+572312.7       &  \multicolumn{2}{c}{\nodata}  & 2.686 &     \multicolumn{2}{c}{\nodata}        &  5.7 & 1.7 &    \multicolumn{2}{c}{\nodata}      & 3.3 & 0.8 &  46.3 & 10.2 \\[-6pt] 
\enddata

\tablecomments{MAMBO $1200\,\mu$m fluxes are from \citet{Greve2004}, upper limits (bracketed fluxes) are from T. Greve (2006, private communication), Bolocam $1100\,\mu$m fluxes are from \citet{Laurent2006}, and 850\,$\mu$m SCUBA and $1.4\,$GHz fluxes are taken from \citet{Chapman2005}. \citet{BiggsIvison2006} provide alternative radio fluxes for many of the listed objects. The bracketed IDs indicate sources that have not been observed by the authors within the context of this paper, but archived data are used in the analysis.}
\tablenotetext{a}{SHARC-2 detection offsets are with respect to the published radio positions.}
\tablenotetext{b}{Both of these sources are likely contributors to the observed $1100 \,\mu$m flux. The Bolocam data are ignored in the analysis.}
\tablenotetext{c}{The alternative radio fluxes from \citet{BiggsIvison2006} are significantly different from or more accurate than the values listed here.}
\tablenotetext{d}{SHARC-2 $350\,\mu$m fluxes from \citet{Laurent2006}.}
\tablenotetext{e}{\citet{Chapman2005}: ``These SMGs have double radio identifications, one lying at the tabulated redshift and a second lying at z $<$ 0.5.'' The higher redshift source is assumed to be the dominant contributor.}
\tablenotetext{f}{\citet{Chapman2005}: ``These SMGs have double radio identifications, both confirmed to lie at the same redshift.''}

\end{deluxetable*}

The observations were carried out during eight separate observing runs between 2002 November and 2005 April, in excellent weather ($\tau_{225\,{\rm GHz}} < 0.06$), reaching $1\,\sigma$ depths of 5--9\,mJy in 2--4 hours of integration in $14$ small fields ($\simeq 2.5 \times 1$ arcmin$^2$) around the targeted sources. Our scanning strategy was to modulate the telescope pointing with a small-amplitude (15''--20'') nonconnecting Lissajous pattern within the limits of reasonable telescope acceleration (with typical periods of 10--20 s). This pattern was chosen to provide fast, two-dimensional, recurring but non-closed patterns with crossing paths---all of which are essential to allow the separation of the superposed source, atmospheric, and instrumental signals. For the observations since 2003 February, we have taken advantage of the CSO's Dish Surface Optimization System \citep{Leong2005} to improve beam shapes and efficiencies, and hence sensitivities, at all elevations.

Pointing corrections were retroactively applied to the observations by a careful fit to the pointing data taken during each run. In addition, our preliminary source identifications revealed a small, albeit significant ($\simeq 3''$) systematic pointing shift in the negative R.A.\ direction. All maps were realigned accordingly. The reconstructed pointing is accurate to $3''\!\!\!.5$ rms, in good agreement with CSO specifications. 

The data were processed using the CRUSH\footnote{See {http://www.submm.caltech.edu/~sharc/crush}} software package, developed at Caltech \citep{Kovacs2006}, which models total power detector signals using an iterated sequence of maximum likelihood estimators. The FWHM point-spread functions of the final maps (Fig.~\ref{fig:maps}) are approximately $12''$ after optimal filtering to search for beam-sized ($\simeq 9''$) features in order to yield maximal signal-to-noise ratios on point sources. This degradation of the image resolution by approximately $\sqrt{2}$, from the nominal instrumental beam width of $8''\!\!\!.5$, is because the fitting for beam-sized features is equivalent to convolving the image with the slightly wider effective beam of around $9''$, which accounts for the smearing of the nominal beam by pointing and focus variations on the long integrations. 

Calibration was performed primarily against planets and asteroids, when available, or using stable galactic continuum sources and Arp 220.\footnote{See {http://www.submm.caltech.edu/~sharc/analysis/calibration.htm}} Trends in aperture efficiency, especially with elevation, were estimated and were taken into account when calibrating our science targets. The systematic flux filtering effect of the aggressive reduction parameters used in CRUSH to subtract noise signals and reach maximal image depths, was carefully estimated for the case of point sources, and the appropriate corrections were applied to the images. The final calibration is expected with high confidence to be more accurate than $15\%$, with systematic effects anticipated to be less.

The maps produced by CRUSH reveal Gaussian noise profiles with an anticipated tail at positive fluxes. Hence, in searching for $350\,\mu$m counterparts around the published radio positions, we calculate detection thresholds based on the effective number of beams  $N_{\rm beam} \approx 1+A/A_{\rm beam}$ \citep{Kovacs2006} inside a detection area $A$ in the smoothed image. A detection confidence $C \approx 1$ is reached at the significance level S/N, at which 
\begin{equation}
1 - C \approx \frac {N_{\rm beam}}{\sqrt{2 \pi}} \int_{\rm S/N}^{\infty} e^{- 1/2 x^2} \, dx .
\label{eq:SNR-det}
\end{equation}

As for the appropriate search radius, the probability that the peak of the smoothed image falls at some radius away from the true position is essentially the likelihood that the smoothed noise level at that radius is sufficient to make up for the deficit in the underlying signal. Therefore, for the case of Gaussian noise profiles, we expect the detection to lie inside of a $2\,\sigma$ noise peak, and therefore within a maximal radius given by the condition ${\rm S/N} = 2 / [1 - \exp (- r_{\rm max}^2 / 2\sigma_{\rm beam}^2)]$ for a Gaussian beam profile with $\sigma_{\rm beam}$ spread at an underlying signal-to-noise ratio of S/N. To this we must add the appropriate pointing tolerance (at $2\,\sigma$ level) in quadrature, arriving at an expression for the detection radius around the actual position as a function of detection significance, that is,
\begin{equation}
r_{\rm max}^2 = 4\,\sigma_{\rm pointing}^2 -2\,\sigma_{\rm beam}^2 \ln \left( 1 - \frac{2}{\rm S/N} \right).
\label{eq:R-det}
\end{equation}
Note that this expression simplifies to $r_{\rm max} \rightarrow 2\sigma_{\rm beam}(\rm S/N)^{- 1/2}$ for the case of ${\rm S/N} \gg 1$ and negligible $\sigma_{\rm pointing}$. 

The combination of ${\rm S/N} = 2.30$ and $r_{\rm max} = 10''\!\!\!.4$ simultaneously satisfy both constraints (eqs.~[\ref{eq:SNR-det}] and [\ref{eq:R-det}]) at $C=95\%$ confidence level for a $\sigma_{\rm beam} = 9''$ effective beam and $\sigma_{\rm pointing} = 3''\!\!\!.5$ pointing rms. Potential candidates thus identified are subsequently verified to lie within the expected distance from their respective radio positions. The resulting identifications are summarized in Table~\ref{tab:fluxes}. When counterparts were not found, the peak measurement values inside the search area are reported.

The sources, collectively, are much smaller than the SHARC-2 beam (i.e., $d \ll 9''$), as neither (1) fitting larger ($12''$) beams or (2) filtering extended structures produces systematically different fluxes for the sample as a whole. Source extents typically $\lesssim 30\,$kpc are therefore implied. While the partial resolution of a few objects cannot be excluded, the peak fluxes of the optimally filtered images are expected to be generally accurate measures of the total integrated flux for the objects concerned.  

\section{Spectral Energy Distributions of SMG\lowercase{s}}

We fitted the SHARC-2 $350\,\mu$m and SCUBA $850\,\mu$m fluxes, combined with Bolocam $1100\,\mu$m \citep{Laurent2005} and MAMBO $1200\,\mu$m \citep{Greve2004} data when available, with single-temperature, optically thin greybody models of the form $S(\nu,T) \propto \kappa(\nu) B(\nu,T)$, where $\kappa(\nu) \propto \nu^{\beta}$ is an approximation for the full emissivity term $(1-\exp [(\nu/\nu_0)^\beta])$ for $\nu_{\rm obs} \ll \nu_0$. Alternative SED models incorporating the full optical depth, or a distribution of temperatures and power-law Wien tails, did not provide a better description of the data. Specifically, the flattening of the Rayleigh-Jeans slope due to optical depths approaching unity is not detectable with the $\simeq 10\%$ uncertain relative calibration of the bands, while the Wien side of the spectra is not sampled by the observations. More complex SED models, e.g.\ the two-temperature model used by \citet{DunneEales2001}, were not considered, since these require a greater number of parameters than can be determined from the few, often just two, photometric data points available for the typical SMG. 

\begin{figure}
\includegraphics[width=\columnwidth]{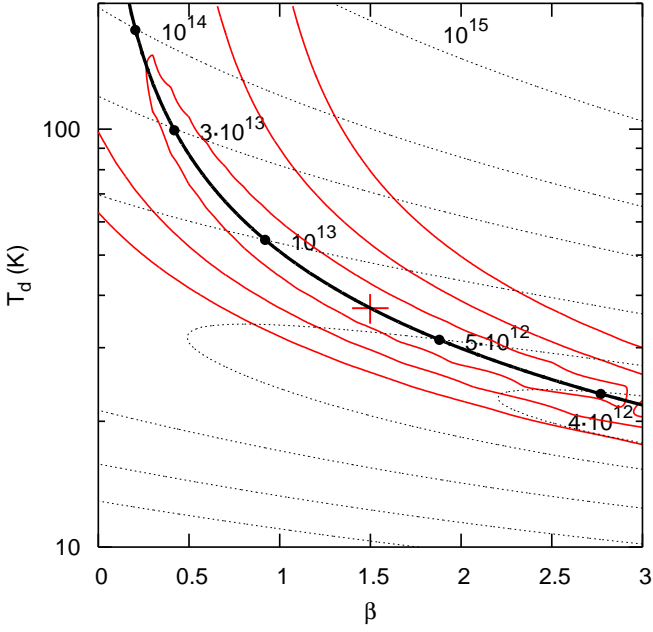}
\caption{
Characteristic dust temperature $T_d$ and effective emissivity index $\beta$ are highly correlated parameters of the SED fit for the typical SMG, even when three photometric data points allow simultaneous fitting of both $T_d$ and $\beta$ (source 13 from Table~\ref{tab:derivatives}, shown with solid 1,2 and 3$\,\sigma$ contours). However, by assuming $\beta=1.5$, we can accurately constrain luminosities ({\it dotted contours}) and temperatures, which are listed in  Tables~\ref{tab:derivatives} and \ref{tab:best-derivatives}. The derived properties may be scaled to other $\beta$ values of the reader's preference using the relations of eq.~(\ref{eq:scaling}) ({\it thick solid line}).
}
\label{fig:T-beta}
\end{figure}

SED models, whether incorporating an emissivity slope, full optical depth or multiple temperature components, are simple parametrizations of complex underlying spectral distributions, produced by a range of dust properties inside the targeted galaxies. Therefore, $T$ and $\beta$ of the model describe not the physical temperature and emissivity of every single dust grain, but provide an effective characterization of dust emission in a galaxy as a whole. In this sense, the characteristic $\beta$-values, expected in the range 1--2, reflect both the underlying grain emissivities and a distribution of physical temperatures within the observed objects. The derived $T$-values provide an effective comparison of the characteristic dust temperatures among objects with equivalent SED parametrization.

Simultaneous fitting of $T$ and $\beta$ requires at least three photometric data points (while many SMGs have only two), and even when permitted these are highly correlated parameters of the fit for the typical SMG (Fig.~\ref{fig:T-beta}). Therefore, we initially assume $\beta = 1.5$ for the SED fit, as it provides good characterization of actively star-forming environments, both in accurately modeled Galactic clouds \citep{Dupac2003} and galaxies of the local universe \citep{DunneEales2001}, and is broadly consistent with laboratory measurements on carbite and silicate grains \citep{Agladze1996}. The properties, thus derived, may be scaled to the $\beta$-values of the reader's choosing, via the deduced spectral indices $\varepsilon$ and $\lambda$ listed in the Tables~\ref{tab:derivatives} and \ref{tab:best-derivatives}, as
\begin{equation}
\label{eq:scaling}
T_d      \propto \beta^{- \varepsilon},~
L_{FIR}  \propto \beta^{- \lambda},~ 
M_d      \propto \beta^{\varepsilon (4+\beta) - \lambda}.
\end{equation}

\begin{deluxetable}{lr@{~$\,\pm$}lr@{~$\,\pm$}lr@{~$\,\pm$}lcc}[t]
\tablewidth{\columnwidth}
\tablecolumns{11}
\tablecaption{Far-Infrared Properties of SMGs from SHARC-2 Data\label{tab:derivatives}}
\tablehead{\colhead{}   & \multicolumn{2}{c}{$T_d$} & \multicolumn{2}{c}{$\log L_{FIR}$}  & \multicolumn{2}{c}{} & \colhead{} & \colhead{} \\ 
  \colhead{ID} & \multicolumn{2}{c}{(K)}   & \multicolumn{2}{c}{($L_\odot$)} & \multicolumn{2}{c}{$q_L$} & \colhead{$\varepsilon$} & \colhead{$\lambda$}}
\startdata
  1  & 43.3&18.7 & 12.83&0.64 & 2.39&0.64 & 1.36 & 2.94 \\
  2  & 66.8&26.1 & 13.49&0.57 & 2.31&0.57 & 1.61 & 3.69 \\
  3  & 21.3&3.8  & 11.95&0.24 & 1.91&0.26 & 0.50 & 0.21 \\ 
  4  & 40.1&5.8  & 12.98&0.19 & 2.33&0.21 & 0.84 & 1.06 \\
  5  & 39.1&4.2  & 13.03&0.14 & 2.39&0.21 & 0.69 & 0.66 \\
  6  & 25.7&4.7  & 12.28&0.18 & 1.71&0.18 & 0.52 & 0.35 \\
  7  & 31.0&5.3  & 12.73&0.15 & 2.19&0.15 & 0.54 & 0.39 \\ 
  8  & 40.9&8.2  & 12.85&0.25 & 2.20&0.26 & 0.57 & 0.46 \\ 
  9  & 21.1&6.3  & 11.95&0.29 & 1.16&0.29 & 0.31 & 0.04 \\ 
 10  & 24.3&6.8  & 11.88&0.35 & 0.79&0.35 & 0.59 & 0.50 \\ 
 11  & 53.4&16.2 & 13.18&0.42 & 2.61&0.43 & 1.38 & 2.88 \\
 12  & 34.5&4.5  & 12.72&0.17 & 1.75&0.17 & 0.73 & 0.77 \\
 13  & 37.3&5.0  & 12.93&0.16 & 2.31&0.18 & 0.77 & 0.91 \\
 14  & 16.8&2.3  & 11.38&0.15 & 2.13&0.22 & 0.65 & 0.53 \\
 15  & 31.5&4.2  & 12.71&0.16 & 2.23&0.21 & 0.65 & 0.57 \\
(17) & 27.4&3.0  & 12.41&0.15 & 2.02&0.16 & 0.61 & 0.44 \\
(19) & 14.2&2.2  & 10.97&0.21 & 1.82&0.23 & 0.58 & 0.41 \\
(20) & 40.6&12.3 & 12.87&0.48 & 2.76&0.50 & 0.99 & 1.61 \\[-6pt]
\enddata
\tablecomments{A summary of the derived far-infrared properties of the observed SMGs. All quantities were derived using an optically thin approximation with $\beta=1.5$. Temperatures and luminosities may be scaled to other $\beta$-values using the relations of eq.~(\ref{eq:scaling}) and the corresponding indices $\varepsilon$ and $\lambda$ listed here. The $q_L$ values derived from the radio data of \citet{BiggsIvison2006} tend to be higher than these by 0.06 on average. }
\end{deluxetable}

\begin{deluxetable}{lr@{~$\,\pm$}lr@{~$\,\pm$}lr@{~$\,\pm$}lcc}[bt]
\tablewidth{\columnwidth}
\tablecolumns{9}
\tablecaption{Properties of SMGs Incorporating Radio Data\label{tab:best-derivatives}}
\tablehead{\colhead{} & \multicolumn{2}{c}{$T_d$} & \multicolumn{2}{c}{$\log L_{FIR}$} & \multicolumn{2}{c}{$\log M_d$} & \colhead{} & \colhead{} \\ 
  \colhead{ID} & \multicolumn{2}{c}{(K)}   & \multicolumn{2}{c}{($L_\odot$)} & \multicolumn{2}{c}{($M_\odot$)} & \colhead{$\varepsilon$} & \colhead{$\lambda$} }

\startdata
  1  & 37.2&3.8 & 12.60&0.12 & 8.64&0.17 & 0.82 & 1.09 \\
  2  & 60.3&6.1 & 13.34&0.12 & 8.22&0.16 & 0.86 & 1.15 \\
  3  & 23.5&2.5 & 12.10&0.15 & 9.24&0.17 & 0.72 & 0.82 \\
  4  & 37.1&3.4 & 12.86&0.11 & 8.91&0.14 & 0.77 & 0.89 \\
  5  & 36.8&3.3 & 12.94&0.12 & 9.01&0.12 & 0.67 & 0.65 \\
  6  & 38.6&7.0 & 12.66&0.13 & 8.61&0.33 & 0.94 & 1.29 \\
  7  & 30.3&3.9 & 12.71&0.09 & 9.24&0.25 & 0.68 & 0.78 \\
  8  & 39.8&4.5 & 12.81&0.12 & 8.69&0.19 & 0.71 & 0.93 \\
  9\tablenotemark{a}  & 21.1&6.3 & 11.95&0.29 & 9.36&0.48 & 0.31 & 0.04 \\
 10\tablenotemark{a}  & 24.3&6.8 & 11.88&0.35 & 8.93&0.37 & 0.59 & 0.50 \\
 11  & 41.5&4.4 & 12.82&0.13 & 8.60&0.16 & 0.80 & 0.99 \\
 12  & 43.2&4.7 & 13.01&0.10 & 9.69&0.17 & 0.96 & 1.40 \\
 13  & 34.8&3.2 & 12.84&0.10 & 9.04&0.14 & 0.73 & 0.83 \\
 14  & 16.8&1.9 & 11.38&0.14 & 9.32&0.19 & 0.72 & 0.74 \\
 15  & 30.7&3.2 & 12.68&0.12 & 9.18&0.16 & 0.70 & 0.75 \\
(17) & 28.6&2.3 & 12.47&0.12 & 9.14&0.13 & 0.75 & 0.85 \\
(19) & 16.4&1.7 & 11.17&0.13 & 9.17&0.16 & 0.81 & 1.02 \\
(20) & 28.6&3.5 & 12.31&0.17 & 8.98&0.15 & 0.76 & 1.02 \\[-6pt] 
\enddata
\tablecomments{Quantities were derived similarly to Table~\ref{tab:derivatives}, except that the radio data are also incorporated with $q_L = 2.14$. The estimate of the dust masses additionally assumes $\kappa_d(850\,\mu{\rm m}) = 0.15\,$m$^2\,$kg$^{- 1}$ for the dust absorbtion efficiency.}
\tablenotetext{a}{These SMGs are identified as being radio-loud. Therefore, the radio data are not used in constraining the far-infrared SEDs.}
\end{deluxetable}

\begin{figure}
\includegraphics[width=\columnwidth]{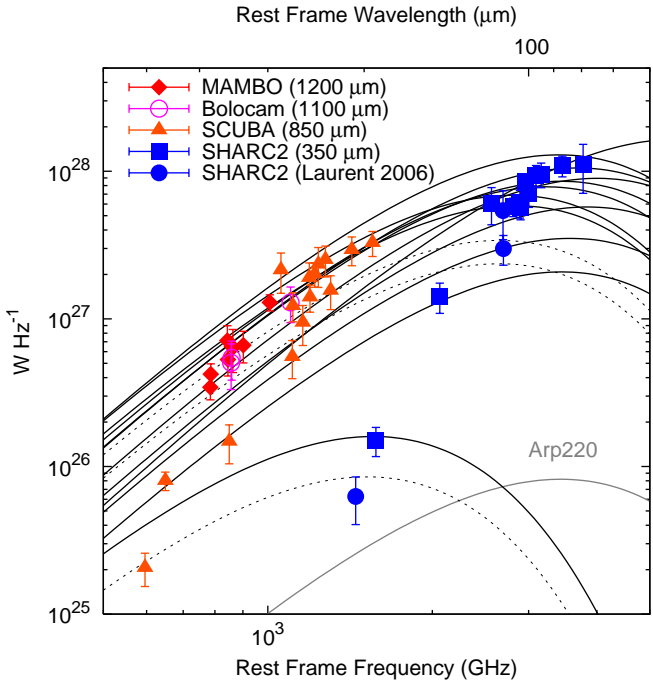}
\caption{
Optically thin greybody SED models were fitted ({\it solid lines}) to the various available observed fluxes and are transformed to the rest frames of the galaxies for comparison. Fits using \citep{Laurent2006} data also shown ({\it dashed lines}). All the spectra peak in the neighbourhood of the redshifted $350\,\mu$m point, providing testimony of the pivotal importance of the SHARC-2 data point in constraining these models. The SEDs shown represent our best fit to the data for sources detected by SHARC-2, use an emissivity index $\beta=1.5$, and incorporate radio fluxes via the far-infrared to radio correlation with $q_L=2.14$ with an assumed 0.12 dex dispersion.
}
\label{fig:SEDs}
\end{figure}

We included $15\%$ calibration uncertainty in addition to the published statistical uncertainties for all submillimeter data. Nevertheless the luminosities are constrained accurately because the SHARC-2 $350\,\mu$m measurement falls near the emission peak for all of these sources (Fig.~\ref{fig:SEDs}).

Flux boosting or Eddington bias \citep{Coppin2005}, in the case of less significant detections, where the selection is in the submillimeter, induce a small bias in the derived SEDs. As the bias will not apply to follow-up measurements, and because the exact quantification of the Eddington bias is non-trivial, requiring a-priori knowledge of the underlying source distributions, we left fluxes uncorrected for its effects. 

For the first time we are able to use multiband photometry to accurately determine the characteristic dust temperatures and far-infrared luminosities for the SMG population. Luminosities (Tables~\ref{tab:derivatives} and \ref{tab:best-derivatives}) are calculated analytically from the fitted grey body model $S(\nu,T)$, following \citet{DeBreuck2003}:
\begin{equation}
\label{eq:Lfir}
L_{FIR} = 4 \pi D_L^2 \Gamma(4+\beta) \zeta(4+\beta) \left( \frac {k T} {h \nu} \right) ^{4+\beta} 
\left( e^{h \nu / k T}-1 \right) \nu S(\nu,T),
\end{equation}
where luminosity distances ($D_L$) were obtained\footnote{See {http://www.astro.ucla.edu/~wright/CosmoCalc.html}} for a $\Lambda$CDM cosmology with $H_0=65\,$km\,s$^{-1}$\,Mpc$^{-1}$, $\Omega_M = 0.3$ and $\Omega_{\Lambda} = 0.7$. The above expression provides the correct SED integral as long as the transition from the optically thin greybody approximation to optically thick blackbody is above the emission peak where the contribution to the total luminosity is negligible. If power-law Wien tails, with spectral slopes of $-\alpha$ \citep{Blain2003}, are assumed instead, the luminosities can be scaled by a constant $\eta_{\beta}(\alpha)$, which is around 1.5 for an Arp220 type template with $\alpha = 2$. More generally, in the range of $\alpha \sim 1.1$--4.0 and $\beta \sim 0$--3, the values of $\eta$ are well approximated (with an rms of $5\%$ or $0.02$ dex) by the empirical formula
\[
\eta_{\beta}(\alpha) \approx (1.44 + 0.07~\beta)(\alpha-1.09)^{-0.42}.
\] 
Similar corrections may be derived for the case of increasing optical depth, with $\eta$ as a function of $h\nu_0/kT$. 

Illuminated dust masses (Table~\ref{tab:best-derivatives}) were also estimated from the SED model $S(\nu, T)$, using \citep{DeBreuck2003},
\[
M_d = \frac {S(\nu, T) D_L^2} {(1+z) \kappa_d(\nu_{\rm rest}) B(\nu_{\rm rest}, T_d) }.
\]
Here, the normalization for the absorption efficiency was assumed to be $\kappa_{850\,\mu{\rm m}} = 0.15\,$m$^2$\,kg$^{-1}$, representing the extrapolated average $125\,\mu$m value of $2.64\pm0.29\,$m$^2$\,kg$^{-1}$ \citep{Dunne2003} from various models 
by assuming $\beta$ of 1.5. 
In comparison to the gas masses derived from\footnote{The CO fluxes from \citet{Tacconi2006} were converted into gas masses using the conversion factor $X_{\rm CO}$ of $0.8\,M_\odot\,($K\,km\,s$^{-1}$\,pc$^2)^{-1}$  in \citet{Solomon1997} appropriate for ULIRGs \citep{Downes1998}. The results are therefore consistent with \citet{Greve2005}.} CO measurements \citep{Greve2005,Tacconi2006}, we find an average gas-to-dust ratio of $54^{+14}_{- 11}\:\bigl( \kappa_{850\,\mu{\rm m}} / 0.15\,$m$^2$\,kg$^{-1}\bigr)$, resembling the ratios seen in nuclear regions of local galaxies by \citet{Seaquist2004}, who assume a $\kappa_d X_{\rm CO}$ product comparable to ours. The individual measurements indicate an intrinsic spread of $\simeq 40$\% around the mean value. The low ratios may be interpreted as an indication for the relative prevalence of dust in SMGs over the local population, which typically exhibit Milky Way like gas-to-dust ratios around $120$ \citep{Stevens2005} or, alternatively, that absorption in SMGs is more efficient with $\kappa_{850\,\mu{\rm m}} \approx 0.33\,$m$^2$\,kg$^{-1}$.

Our estimates of the dust temperatures, and hence luminosities, are systematically lower, than those anticipated based on the $850\,\mu$m and $1.4\,$GHz fluxes alone by \citet{Chapman2005}, who overestimate these quantities by $13\%$ and a factor of $\simeq 2$, respectively, when assuming the local far-infrared to radio correlation. This discrepancy can be fully reconciled if a different constant of correlation is assumed for SMGs (see Section~\ref{sec:correlation}).  

\looseness-1
We confirm that the SMG population is dominated by extremely luminous (several times $10^{12}\,L_{\odot}$) systems with $\simeq 10^9\,M_{\odot}$ of heated dust, and characteristic $35$\,K dust temperatures typical to actively star forming ULIRGs. As anticipated, these objects resemble the nearby archetypal ULIRG, Arp 220 (with $T_d \approx 37\,$K, similarly obtained), except that they are about 7 times more luminous on average. 

\subsection{Cold, Quiescent SMG\lowercase{s}?}

In addition, there appear to be several cooler ($T_d \lesssim 25\,$K), less luminous ($10^{11}$--$10^{12}\,L_{\odot}$) objects, albeit with comparable dust masses (few times $10^9\,M_{\odot}$) present in the population (sources 14 and 19, and possibly 3, 9, and 10 in Tables~\ref{tab:derivatives} and \ref{tab:best-derivatives}). While these resemble the Milky Way in temperatures, and hence probably in star formation densities, they are tens of times more luminous than the Galaxy.

This combination of extreme dust masses yet surprisingly low relative star formation rates, indicated by the lesser dust heating, allude to possibly incorrect low-redshift identifications. Should these galaxies lie at the higher, more typical redshifts of the SMG population, their temperatures and luminosities would closely resemble those of the hotter population. However, while lensing of distant SMGs by massive foreground galaxies at the measured redshifts, is conceivable \citep{Blain1999a, Chapman2002}, these should be rare, and unlikely to account for all cold SMGs.

\begin{figure}
\includegraphics[width=\columnwidth]{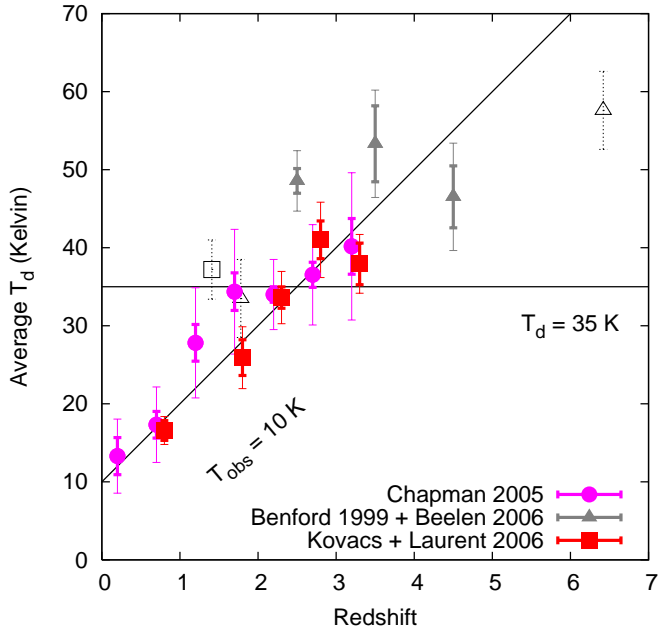}
\caption{
Median dust temperatures vs.\ redshift. Data are binned for this paper ({\it squares}), for the \citet{Chapman2005} sample ({\it circles}) with temperatures estimates incorporating the radio data, and for dusty quasars ({\it triangles}). One-sigma error bars were estimated from the scatter inside the bins and the uncertainties of dust temperatures where available ({\it thick bars}). Dispersion within bins are also shown ({\it thin bars}), while dotted error bars indicate measurement error for single datum inside a bin. A $T_d$ of $35\pm3\,$K may be applicable for $z \sim 1.5$--3.5, which, if true, can provide photometric redshifts in that range. However, observing frame temperatures of around $10\,$K appear to fit the SMG data somewhat better than constant $T_d$ in the rest frame. This similarity of the SEDs in the observing frame may owe to selection effects.
}
\label{fig:T-z}
\end{figure}

Alternatively, these seemingly normal type galaxies could, potentially, be remnants to the more remote, hotter population of SMGs, once the short lived starbursting activity subsides. Dust could persist and remain detectable in the more proximate universe, provided that radiation from the rapidly growing black holes \citep{Borys2005, Alexander2005a} does not disperse dust entirely, e.g. dusty QSOs of \citet{Benford1999} and \citet{Beelen2006}, or if the dust can later reassemble to form large disk galaxies, like source $14$, which is consistent with the large spiral disk galaxy seen in {\it Hubble Space Telescope} images \citep{Almaini2006, Borys2006}, or the cold dust mergers suggested by the chain of $16~\mu$m emission seen for source $19$. This connection between the hot and cold SMGs is further hinted at by the apparent trend of dust temperatures increasing with redshift (Fig.~\ref{fig:T-z}) even without the low-redshift data. We, therefore, find the existence of massive cold SMGs plausible.

\section{The Far-IR/Radio Correlation at High Redshifts}
\label{sec:correlation}

One of the most valuable uses of the new measurements is to test the local far-infrared to radio correlation at the redshifts of SMGs ($z \sim 1$--3). The simple extension of the correlation into the distant universe was suggested by earlier studies based on a combination of $1.4\,$GHz radio with {\it Infrared Space Observatory (ISO)} $15\,\mu$m \citep{Garrett2002} and {\it Spitzer} 24 and 70\,$\mu$m observations \citep{Appleton2004}, both of which take advantage of SED template fitting to extrapolate fluxes into the far-infrared regime. More recently, \citet{Beelen2006} hints at the validity of the correlation in the case of a handful of distant ($z\sim 2$--6) quasars with SEDs well constrained directly from far-infrared measurements. 

A quantitative treatment of the correlation \citep{Helou1985} has been formulated in terms of the {\it Infrared Astronomical Satellite (IRAS)} 60 and 100\,$\mu$m fluxes, expressed in the rest frame as
\begin{equation}
\label{eq:q}
q = \log \left( \frac {\rm FIR} {3.75 \times 10^{12}\,{\rm W}\,{\rm m}^{- 2}} \right) - \log \left( \frac {S_{\rm 1.4\, GHz}} {{\rm W}\,{\rm m}^{- 2}\,{\rm Hz}^{- 1}} \right),
\end{equation}
where the far-infrared parameter ${\rm FIR} = 1.26 \times 10^{- 14} (2.58\,S_{60\,\mu{\rm m}} + S_{100\,\mu{\rm m}})\,$W\,m$^{- 2}$ and the fluxes $S_{60\,\mu{\rm m}}$ and $S_{100\,\mu{\rm m}}$ are in Jy. The quantity essentially estimates the flux in a wide band centered at $80\,\mu$m and is a good tracer of the far-infrared luminosity for temperatures in the range of 20--80\,K and emissivity indices $\beta \sim 0$--2 \citep{Helou1988}.

Perhaps a better measure of the correlation is obtained by directly comparing luminosities. Therefore, we propose to use
\begin{equation}
\label{eq:qL}
q_L = \log \left( \frac {L_{\rm FIR}} {[4.52\,{\rm THz}] \, L_{1.4\,{\rm GHz}}} \right).
\end{equation}
Here the normalization frequency $4.52\,$THz has been derived for the adopted greybody model\footnote{If SEDs include power law Wien tails instead, the normalizing frequency should be appropriately scaled by $\eta_{\beta}(\alpha)$. Similar adjustements to the normalization can be derived for any other SED model, hence the shape of the particular SED model used has no effect on the correlation parameter $q_L$ otherwise} such that $q_L \rightarrow q$ for $T=40\,$K and $\beta = 1.5$, thus approximating the original definition of $q$ while extending its usefulness beyond the original restrictions in temperature and emissivity. Radio luminosities, in turn, are calculated as
\begin{equation}
\label{eq:Lrad}
L_{1.4\,{\rm GHz}} =  4 \pi D_L^2  S_{1.4\,{\rm GHz}} (1+z)^{\alpha-1},
\end{equation}
which includes a bandwidth compression by $(1+z)^{-1}$ and a $K$-correction $(1+z)^{\alpha}$ to rest frame $1.4\,$GHz. We assume a synchrotron power law of $S \propto \nu^{-\alpha}$, with the spectral index $\alpha$ for which a typical value of $0.7$ \citep{Condon1992} of nonthermal radio sources is assumed for the calculations.\footnote{The assumption of the radio spectral index $\alpha$ is not critical. Indices different by $\delta \alpha$ may produce a small bias in $q$ on the order of $\delta q \approx 0.5 \delta \alpha$ in the redshift range of $z \sim 1$--3 of SMGs, with an undetectable redshift dependence as $\delta q \ll \sigma_q$.}

\begin{figure}[t]
\includegraphics[width=\columnwidth]{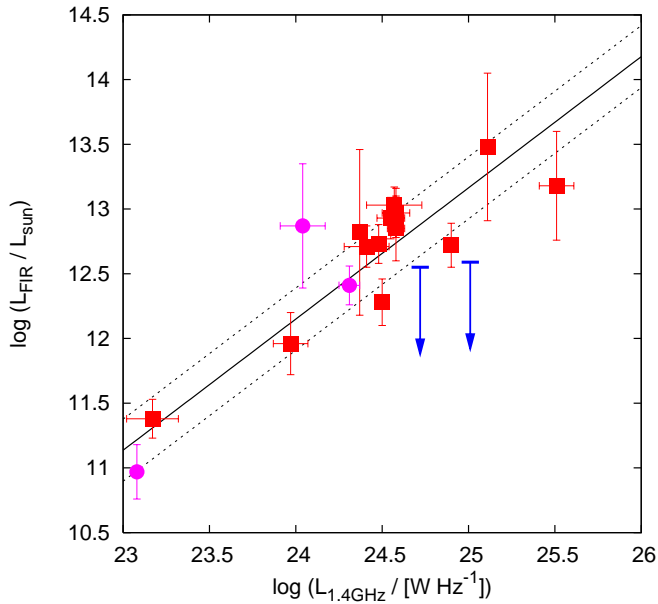}
\caption{
Radio to far-infrared correlation for SMGs. Galaxies observed by the authors shown with squares, data from \citet{Laurent2006} with circles. Two-sigma upper limits are indicated when appropriate ({\it arrows}). Far-infrared luminosities were calculated exclusively based on the available submillimeter measurements, with $\beta=1.5$. The best-fit model ({\it solid line}) reveals no deviation from linearity in the relationship. The deduced 2$\,\sigma$ instrinsic scatters around the model are also shown ({\it dotted lines}). 
}
\label{fig:radio-fir}
\end{figure}

\begin{figure*}[!bt]
\includegraphics[width=0.5\textwidth]{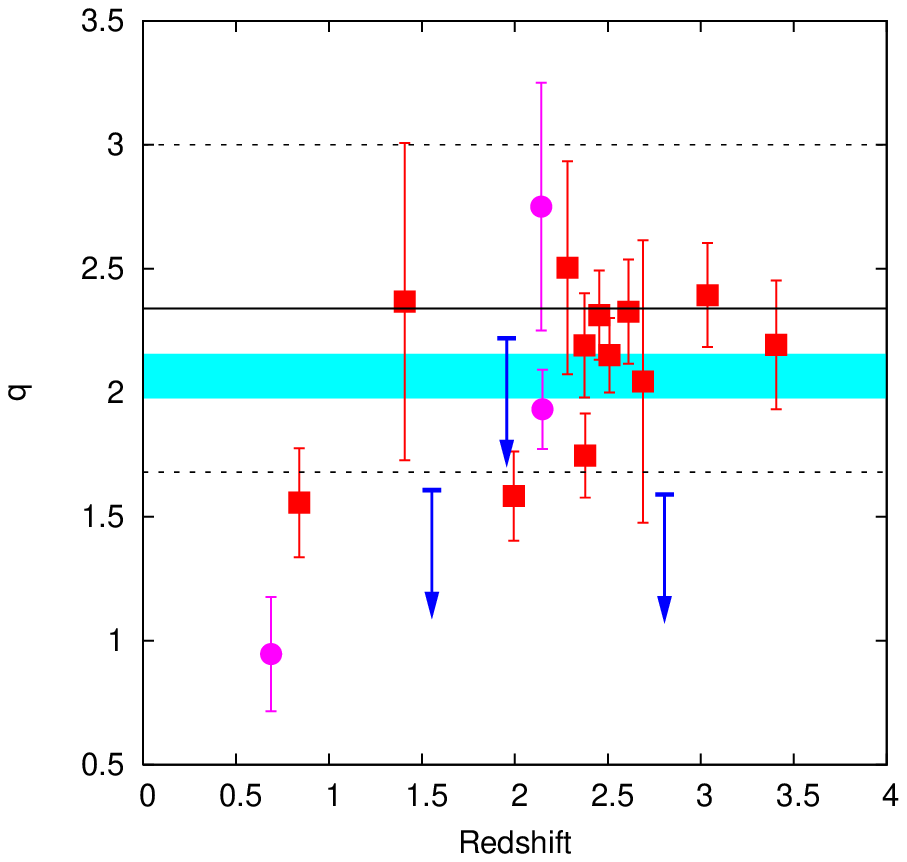}
\includegraphics[width=0.5\textwidth]{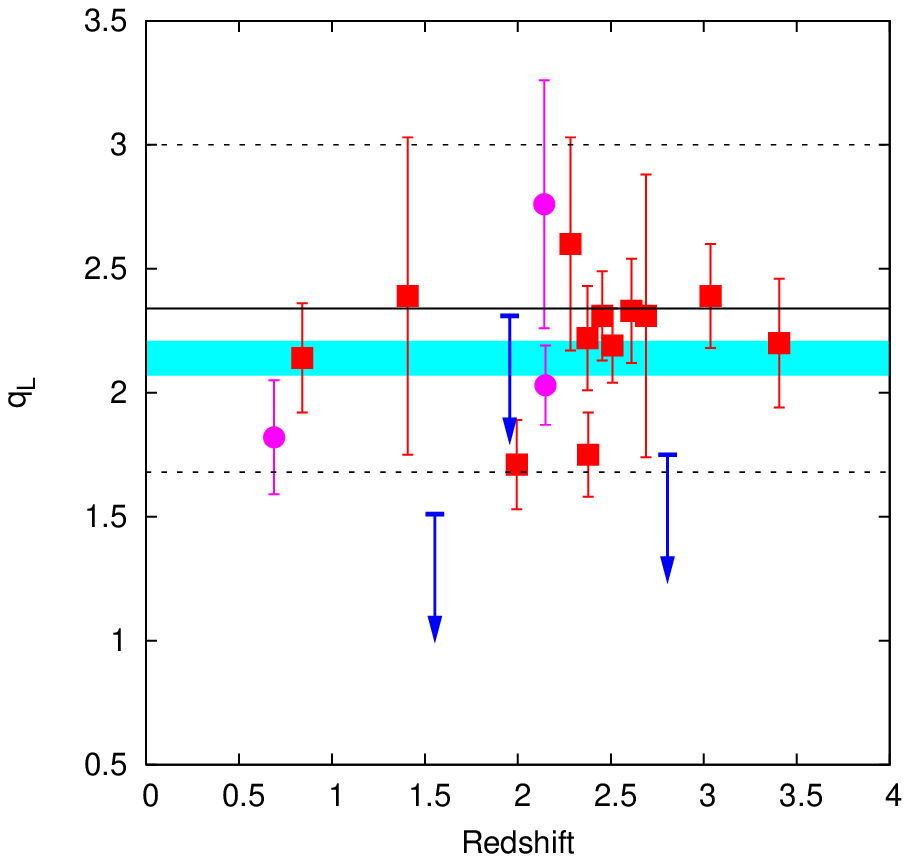}
\caption[The Radio ot Far-Infrared Correlation]{({\it a}) Original definition of $q$ from \citet{Helou1985}, and ({\it b}) the modified, luminosity-based definition $q_L$ as a function of redshift. Two-sigma upper limits are indicated as appropriate ({\it arrows}). Data from this paper are shown with squares, while those from \citet{Laurent2006} are shown as dots. The observed SMG's seem to consistently fall below the locally observed mean of $2.34$ \citep{Yun2001} for local bright {\it IRAS} sources ({\it solid line}) whose 2$\,\sigma$ scatters are also indicated ({\it dashed lines}). The values were all derived assuming $\beta=1.5$. The derived distribution mean values of the non-radio-loud subsample ($q > 1$) are also shown ({\it shaded bands}). 
}
\label{fig:q-vs-z}
\end{figure*}

We confirm that the far-infrared to radio correlation appears to hold throughout the redshift range of $z \sim 1$--3 (Figs.~\ref{fig:radio-fir} and \ref{fig:q-vs-z}), with the notable exceptions of sources 9 and 10, which likely host radio loud AGNs\footnote{Source 9 has been characterised as an AGN by \citet{Chapman2005}}. The detections reveal $\overline{q} \approx 2.07 \pm 0.09$ and $\overline{q}_L = 2.12 \pm 0.07$ with intrinsic spreads estimated at $\sigma_q \approx 0.21$ and $\sigma_{q_L} \approx 0.12$ around the respective mean values. The alternative radio fluxes of \citet{BiggsIvison2006} provide $q$ values that are higher by $0.06$ on average and scatter slightly more with $\sigma_q \approx 0.30$ and $\sigma_{q_L} \approx 0.23$. 

The reduced dispersion in $q_L$ vs.\ $q$ and the somewhat peculiar case of the two cold lower redshift objects (14 and 19), whose derived dust temperature of only $16\,$K fall outside the range where FIR traces luminosity (eq.~[\ref{eq:q}]) highlight the practicality of the luminosity-based $q_L$-measure over the original {\it IRAS}-based $q$. The derived mean values are significantly less than the locally observed median values of $2.75\pm0.03$ for local spirals and starbusts galaxies from the {\it IRAS} Faint Source Catalog \citep{CondonBroderick1991}, $2.34\pm0.01$ for radio-identified, flux-limited ($S_{60\,\mu{\rm m}} > 2\,$Jy) {\it IRAS} sources \citep{Yun2001}, or $\simeq 2.3$ for a range of normal galaxies \citep{Condon1992} that include spirals and irregulars, E/S0 types, {\it IRAS} and radio-selected samples. The corresponding local scatters are also larger, with $\sigma_q = 0.14$, $\sigma_q = 0.33$, and $\sigma_q \lesssim 0.2$, respectively.

Moreover, we find the relationship to be linear, within measurement uncertainty, over nearly three decades of luminosities of the observed SMGs, with a best-fit FIR to radio luminosity index, $d\log L_{\rm FIR} / d\log L_{1.4\,{\rm GHz}}$, of $1.02\pm0.12$. Nonetheless, small levels of nonlinearity, like those discussed by \citet{Fitt1988, Cox1988, Devereux1989} and \citet{Condon1991} remain possible.

The low $q$-values, the tightness of the correlation, the observed linearity, and the typically warm dust temperatures all point to dust heating that is dominated by high-mass ($>8\,M_\odot$) stars. In the two-component model of \citet{Yun2001}, short-lived ($<10^7\,$yr) high-mass stars fuel the ``hot'' thermal component that is strongly coupled to the nonthermal radio via Type II supernovae, whereas the ``cold'' component heated by a different population of lower mass stars is expected to produce a different correlation with a weaker coupling. They estimate $q_1=2.3$ and $q_2 > 4$ for the components. As most galaxies contain both components, intermediate $q$-values are typically observed. The low $q$'s that characterize SMGs are, therefore, indicative of the predominance of high-mass star formation in the dust heating. Moreover, the SMG result forces the revision of the ``hot'' component correlation to $q_1 \lesssim 2.1$, or lower still (by about $0.16$), if steeper radio spectral slopes of $\alpha \simeq 1$ are assumed for these very luminous objects \citep{Yun2001}.

The presence of radio-quiet AGNs would contribute to the total far-infrared emission, biasing towards higher values of $q$. Radio-loud AGNs, on the other hand, would produce low $q$-values, albeit to wildly varying extent, which should manifest as an increased spread in the correlation. The tightness of the observed correlation, therefore, confirms that radio-loud AGNs are rare, while the low $q$-values, otherwise observed, indicate that the AGN contribution to the far-infared luminosity is small when compared to heating from star formation activity in SMGs. This is consistent with \citet{Alexander2005a, Alexander2005b}, who used ultra-deep {\it Chandra} X-ray data to find that the AGNs, present in at least 75\% of SMGs, contribute little to the total bolometric luminosity of their hosts. Dust heating, therefore, is assumed to be star-formation-dominated. 

Last but not least, the low $q$-values observed may partially arise from selection bias, as our initial sample would miss SMGs with low radio fluxes, i.e., sources with higher $q$-values, in the secondary selection required for redshift identification. However, faint radio sources with undetectable submillimeter counterparts, representing a hypothetised hot extension to the SMG population, may be more typical \citep{Chapman2004, Blain2004}, and thus selection bias could be reversed, missing the low $q$-values instead.

\section{Dust Emissivity Index}
 
It is possible to further constrain the dust properties, e.g.\ by measuring the effective emissivity index $\beta$. In principle, this may be attempted on an individual basis, since several galaxies have MAMBO \citep{Greve2004} and/or Bolocam \citep{Laurent2005} detections at longer wavelengths, providing justification for a three-parameter fit. In practice, however, such an approach is unlikely to yield robust quantities for individual galaxies (see Fig.~\ref{fig:T-beta}) owing to the proximity of the observed bands and the modest significance of the detections. Therefore, we aimed to fit a single emissivity slope $\beta$ for the entire sample, hoping to provide a better constrained, ensemble-averaged, dust emissivity index. Moreover, we extend our sample to \citet{Chapman2005} SMGs with multiband submillimeter data that were not observed in this paper. MAMBO associations \citep{Greve2004} and the $350\,\mu$m follow-up of Bolocam and candidates \citep{Laurent2006} provide five extra sources. 

We analyse the data consisting of detections only separately from the data including measurement peaks inside the search area for nondetections. While the first set may appear more robust, it could be more susceptible to selection biases due to the rigorous detection requirement. The larger set, containing upper limits, is thus likely to be less biased of the two and could provide a better characterisation of the whole SMG population.  

We have used a modified, nested version of the downhill simplex method \citep{Press1986} to perform the fit, yielding $\beta = 2.42 \pm 0.71$, a value that is too uncertain to distinguish between the expected values in the range of 1--2. However, the demonstrated validity of the linear far-infrared to radio correlation for the SMG sample allows us to incorporate the radio fluxes to constrain the total integrated far-infrared luminosities and improve the quality of the fit. Here we assumed $0.12$ dex as the intrinsic spread in $q_L$, producing the expected model deviation with $\hat{\chi}^2 \rightarrow 1$, and consistent with the scatter observed for local spirals and starbursts \citep{Condon1991}.

To avoid radio-loud objects biasing our analysis, we ignored sources 9 and 10 due to their a priori low $q_L$-values. At the same time, we utilized a modified $\chi$ deviation definition to allow some of the sources to become ``radio-loud'' in the fit, i.e.\ to significantly deviate from the correlation in the lower $q_L$ direction. Specifically, when the derived $q_L$ value for a source is more than $2\,\sigma$ below the average $\overline{q}_L$ of the fit, we use a relaxed deviation with $\sigma_{q_L}' = 2$, more characteristic of radio loud objects, while keeping $\chi^2$ continuous over the entire parameter space. The fraction of normal sources falsely identified as ``radio-loud'' by this method is expected to be around $2\%$, and therefore the effects of mischaracterizations are expected to remain limited.

With the local median $q$ value of $2.34$ we obtain $\beta = 0.81 \pm 0.19$ for the detections and $\beta = 1.00 \pm 0.19$ when all the available information is used in the analysis. Both these results imply an emissivity index significantly less than the usually quoted values around $1.5$. The interpretation, however, relies entirely on the assumption that the local radio-far-infrared correlation holds unaltered for higher redshifts. 

\begin{figure}[!b]
\includegraphics[width=\columnwidth]{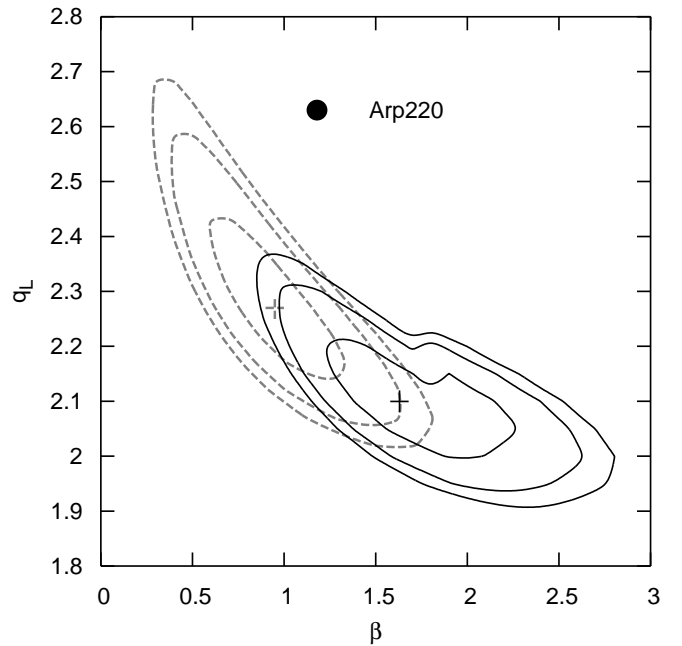}
\caption[$\beta - q$ Contours]{
Likelihood contours (at 68\%, 90\%, and 95\% confidence) for the dust emissivity $\beta$ and the radio to far-infrared correlation constant $q_L$, using only the available detections ({\it dashed curves}) and all information, including nondetections ({\it solid curves}). The best-fit locii of both fits are indicated by crosses. The more inclusive set is expected to be a more reliable indicator of SMG properties, as it is less affected by the selection biases that are introduced by requiring detections in all bands. A similar fit to Arp 220 is also indicated ({\it black dot}).
}
\label{fig:b-q-contours}
\end{figure}

For a more rigorous treatment, we obtained confidence contours for the likely combinations of both $\beta$ and $q_L$ (Fig.~\ref{fig:b-q-contours})\placefigure{fig:b-q-contours}, hence avoiding a priori assumptions on both of these quantities. The smaller set, with detection requirement imposed, favours somewhat lower $\beta$-values, in combination with ``normal'' $q_L$, with the best fit located at the respective values of $0.95$ and $2.27$, whereas the more inclusive data set tends towards emissivity indices in line with expectations, but with $q_L$ decidedly below the mean in the local universe. The best-fit values of the extended set are at $\beta = 1.63$ and $q_L = 2.10$. Accordingly, we assumed $\beta = 1.5$ and used the corresponding $q_L = 2.14$ (from Fig.~\ref{fig:b-q-contours}) to calculate most likely estimates of the far-infrared emission properties, which are listed in Table~\ref{tab:best-derivatives}.

Systematic biases in the cross-calibration, between the SHARC-2 and the longer wavelength data points, would introduce bias into the estimated emissivity index. Fortunately, with expected $10\%$ relative calibration error, the resulting bias is only $\delta \beta \simeq 0.08$, and our conclusions on the dust emissivity are nearly unaffected.

\section{Photometric Redshifts}

Obtaining accurate redshifts for distant SMGs has relied on optical measurements, guided by radio or optical associations \citep{Chapman2003, Chapman2005}. This approach has been very successful, providing accurate redshifts for nearly a hundred SMGs to date, but it is unlikely to be practical for compiling much larger samples. Moreover, the identifications involving radio or optical counterparts could introduce selection biases into the samples. For both reasons a ``holy grail'' of large submillimeter surveys has been to obtain redshifts directly from the radio, submillimeter, and mid- to far-infrared (e.g.\ {\it Spitzer}) if possible.

\begin{figure*}
\includegraphics[width=0.5\textwidth]{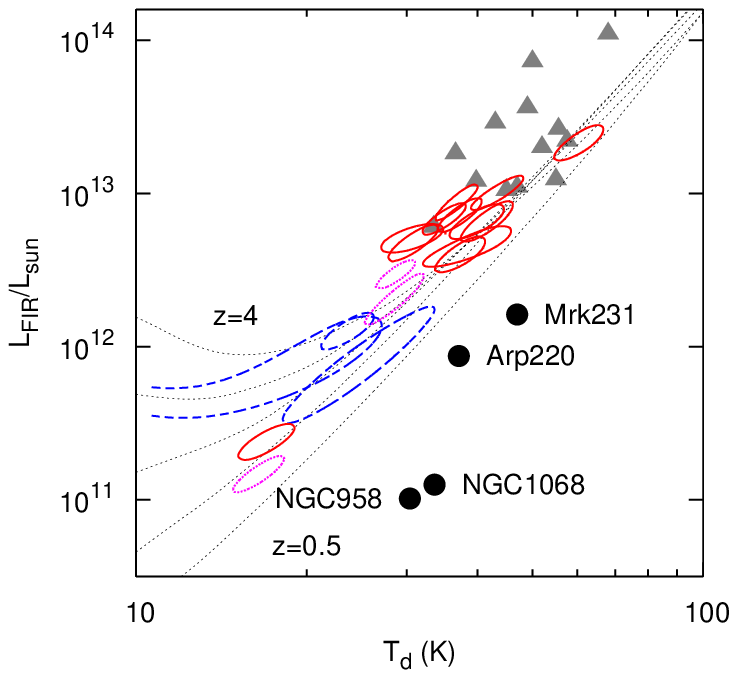}
\includegraphics[width=0.5\textwidth]{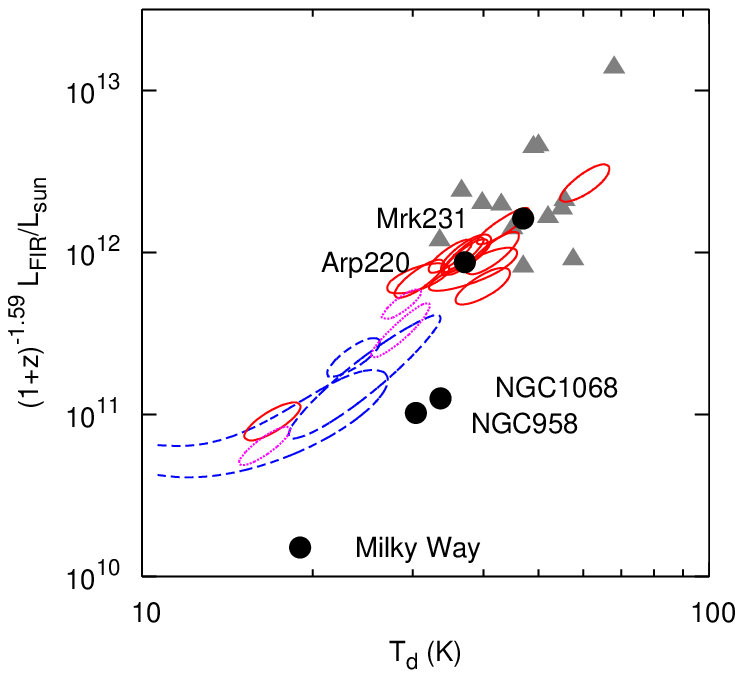}
\caption{
Luminosity-temperature ($L$-$T$) relation. One-sigma likelihood contours for greybody models fitted to submillimeter data (this paper with solid curves, Laurent et al.\ [2006] as dotted curves). Contours are also shown for objects not detected by SHARC-2 ({\it dashed}). The dusty quasars ({\it triangles}) of \citet{Benford1999} and \citet{Beelen2006}, and some well-known objects with accurately constrained SEDs are also shown. The plot highlights the difficulty of measuring the underlying temperature dependence of luminosity, as the two parameters are correlated; most data points are heavily clustered and clearly affected by selection effects. When such a relation is established, however, it can dispute the idea of luminosity-based photometric redshifts. The top graph shows the apparent $L$-$T$ relation and the $850\,\mu$m selection functions ($S_{850\,\mu {\rm m}} > 5\,$mJy) at redshifts of 0.5, 1, 2, 3, and 4. ({\it a}) The observed relation is evidently biased by selection. ({\it b}) A bias- and evolution-free version shows the same relation after the observed redshift evolution (including selection bias) has been accounted for, effectively ``moving'' luminosities into the present era. The case of nearby Arp 220 clearly supports the case of explixit redshift evolution of the $L$-$T$ relationship.
}
\label{fig:L-T}
\end{figure*}

The simplest types of redshift estimators, such as \citet{Carilli1999, Carilli2000a, Carilli2000b}, \citet{Wiklind2003} and \citet{Laurent2006}, assume that all SMGs have intrinsically similar rest-frame temperatures. In essence, redshifts are calculated as $(1+z) \approx T_d/T_{\rm obs}$ \citep{Blain1999} for SEDs characterized by $T_{\rm obs}$ in the observing frame and an assumed representative dust temperature $T_d$ for all objects. The methods only differ in how $T_{\rm obs}$ is estimated from the various available flux data. Only independent, and accurate, measurements of actual dust temperatures for SMGs with known redshifts can test the validity of this approach. We offer this test for the first time.

A glance at Tables~\ref{tab:derivatives} and \ref{tab:best-derivatives} reveals that dust temperatures are distributed around $35\,$K, with some $80\%$ of the data between 25--45\,K. The same conclusion is suggested by the larger \citet{Chapman2005} sample, with the dust temperatures estimated from the SCUBA $850\,\mu$m and radio data using the appropriate far-infrared to radio correlation as discussed above. This majority population of SMGs is consistent, within measurement uncertainty, with a single representative temperature of $34.6 \pm 1.4\,$K in the rest frame, with an estimated 3\,K intrinsic dispersion around the mean, lending some credibility to photometric redshifts derived as
\[
(1+z) \approx \frac{(34.6\pm3.0)\,{\rm K}} {T_{\rm obs}} \left( \frac{1.5}{\beta}\right)^{0.71}
\]
in the redshift the range of $z \sim 1.5$--3.5. However, the observations suggest that these photometric redshift indicators could be wholly inappropriate for as many as $20\%$ of all SMGs, unless some of these outliers have incorrect redshifts. Curiously, the very low $q_L$-values measured for sources 14 and 19 hint at ``hot'' type galaxies \citep{Yun2001}, in apparent contradiction with the cool dust temperatures that are implicated at their low redshifts. This mismatch would be reconciled if the temperatures were typical at the photometric $z$-values of $2.96^{+0.51}_{- 0.43}$ and $1.95^{+0.33}_{- 0.28}$, respectively.

While quasars \citep{Benford1999, Beelen2006} too can be characterised by rest-frame SED templates, with $T_d \approx 48\pm8\,$K\footnote{Temperatures from \citet{Beelen2006} were adjusted for $\beta=1.5$ for consistency.} more or less uniformly across $z \sim 2$--7, they are typically hotter than SMGs, perhaps due to additional dust heating by an AGN. However, quasars, selected from larger volumes than SMGs, could provide the rare extension of the SMG population into ever more extreme luminosities and hotter dust temperatures. If true, the different temperatures that characterize quasars versus the bulk of SMGs invalidates the single representative temperature assumption for all SMGs, already under fire from a minority population of cold SMGs, and therefore may oppose rather than support the applicability of temperature-based photometric redshifts.

Alternatively, the {\em entire} SMGs sample, {\em including the low-redshift data}, is better fit by temperature evolution of $T_d \propto (1+z)$, producing similar temperatures in the observed frame ($T_{\rm obs} \simeq 10\,$K), with an average deviation of $|\hat{\chi}| = 1.40$ versus 2.67 produced by the representative rest-frame temperature model when assuming a 10\% intrinsic scatter in $T_d$. Hence, the temperature range that fits most SMGs could arise from the similar fractional dispersion of the redshift distribution, i.e., objects selected near the median redshift of $2.3$ tend to have temperatures of about $35\,$K. Thus, temperature-based photometric redshifts may not provide additional information beyond a reflection of the median $z$ of the sample. 

A similarity of SEDs in the observed frame, yielding a nearly constant $T_{\rm obs}$ and thus $T_d \sim (1+z)$, could arise simply from selection effects that result from the minimum flux requirement for detections (Fig.~\ref{fig:L-T}), and the small dynamic range (of a factor $\simeq 3$) of flux densities that characterise current SMG samples. As such, selection effects alone could render far-infrared and radio based photometric redshift indicators wholly unsuitable for unbiased flux- and volume-limited samples of SMGs.

\subsection{Luminosity-Temperature Relation}
An alternative, luminosity-based approach exploits the hypothetical relationship between dust temperatures and the luminosities that they fuel, to derive photometric redshifts by comparing the rest-frame relationship to observed temperature indicators and fluxes. Like other scaling relations that characterize galaxies, the $L$-$T$ relationship would, presumably, be a genuine property of galaxies, reflecting the physics of galactic structure and dynamics. While the idea has been drafted by \citet{Yun2002}, \citet{Blain2003}, and \citet{Aretxaga2005a}, the particular implementation of such a method was left awaiting the determination of the actual form of such an $L$-$T$ relation.

We find that the far-infrared luminosity is well approximated by a power law of the form 
\begin{equation}
\label{eq:L-T}
L_{\rm FIR} = \mathcal{L}_0 T_d^\gamma \times (1+z)^{\mu}
\end{equation}
for some exponents $\gamma$ and $\mu$. The expression incorporates the possibility of explicit redshift evolution, approximated by the term $(1+z)^\mu$ chosen for convenience. This independent redshift term is expected to absorb selection effects and allow the determination of the underlying temperature dependence.

The possible combinations of $L_{\rm FIR}$ and $T_d$ pairs are correlated (Fig.~\ref{fig:L-T}), which adds difficulty to constraining the underlying relationship. With this warning in mind, our best estimates of the exponents are $\gamma = 2.76 \pm 0.29$ and $\mu = 1.74\pm0.38$ for all data points and $\gamma = 2.61 \pm 0.30$ with $\mu=1.25 \pm 0.48$ excluding the lower redshift points. The corresponding normalizations, $\mathcal{L}_0$, are $3.4\times10^7\,L_\odot$ ($\pm0.33$ dex) and $1.1\times10^8\,L_\odot$ ($\pm0.45$ dex), respectively. The fits were obtained via a two-variable linear regression of the logarithms with error bars (Table~\ref{tab:best-derivatives}) on both the $\log L$ and $\log T$ values. As most galaxies lie in a narrow temperature and luminosity range, we point out that the derived temperature exponents are constrained by relatively few points, some or all of which could be peculiar, and therefore the true uncertainties of $\gamma$ and $\mu$ are likely underestimated. Nonetheless, these scaling models successfully apply to the quasars of \citet{Benford1999} and \citet{Beelen2006}, and the local ULIRGs Arp 220 and Mrk 231. If we consolidate our estimates with the inclusion of these objects, we obtain $\gamma = 2.82\pm0.29$, $\mu = 1.59 \pm 0.18$, and $\mathcal{L}_0 = 3.6\times 10^7\,L_\odot$ ($\pm 0.46$ dex). This relationship, derived for SMGs, quasars, and ULIRGs, could hint at a possible connection among these types of objects; however, it may not apply to the less active galaxies of the local universe (Fig.~\ref{fig:L-T}).

Assuming the exponents and scaling are properly determined, we may attempt to derive a photometric redshift indicator by comparing the expressions for the far-infrared luminosity from equations (\ref{eq:Lfir}) and (\ref{eq:L-T}) and from the definitions in equations (\ref{eq:qL}) and (\ref{eq:Lrad}). In order to facilitate the derivation of analytic solutions, we approximate the luminosity distance by $D_L \approx D_0(1+z)^\delta$, the effect of which is to simplify all redshift dependence to powers of $(1+z)$. This simplification is reasonable for the study of SMGs (leading to an rms $\delta z / z$ of $\simeq 6\%$) in the redshift range of 0.5--4 with $D_0 = 1.63\,$Gpc and $\delta = 2.07\pm0.03$. We find that the far-infrared luminosity is then proportional to
\begin{equation}
\label{eq:z-dependance}
S_{1.4\,{\rm GHz}} (1+z)^{2\delta + 1-\alpha} \propto T_{\rm obs}^{\gamma}(1+z)^{\gamma+\mu} \propto \frac {S_{\nu} T_{\rm obs}^{4+\beta}} {e^{h\nu / kT}-1}~(1+z)^{2\delta}.  
\end{equation}

Substituting the quantities assumed or determined so far, we find that all three expressions possess similar dependences on $(1+z)$, with exponents of $4.44\pm0.12$, $4.41\pm0.34$, and $4.14\pm0.06$, respectively. As the exponents are indistinguishable within the measurement uncertainties involved, we conclude that, in the redshift range of $z\sim 0.5$--4 of the approximation, it is very unlikely that a meaningful measure of redshift can be deduced in this way---bad news for luminosity-based photometric redshift estimation. While it may simply be an unfortunate coincidence, it is more likely that the inability to determine redshifts in this manner is, once again, the product of the sample selection yielding candidates with similar observed SEDs. With these being nearly identical in the observing frame, photometric redshifts of any kind remain elusive.

We can, nevertheless, derive a true luminosity-based photometric redshift indicator in the low-redshift limit $z \ll 1$, where the luminosity distance takes the asymptotic form $D_L \rightarrow cz/H_0$. In this limit, 
\[
z \approx 1\,{\rm Gpc}~ \frac{H_0}{c} \left(\frac{\mathcal{L}_0 T_d^\gamma}{L_{1\,{\rm Gpc}}}\right)^{1/2},
\]
where $L_{1\,{\rm Gpc}}$ is what the luminosity would be if the galaxy were at a distance of $1\,$Gpc. While this is irrelevant to the SMG sample, it provides a useful expression for the study of the $L$-$T$ relation in the nearby ($z \ll 1$) universe, when values of $z$ are already known.

Comparing the $L$-$T$ relationship (eq.~[\ref{eq:L-T}]) with the expression for the integrated far-infrared luminosity (eq.~[\ref{eq:Lfir}]) we infer that $M_d \propto T_d^{-(4+\beta-\gamma)} (1+z)^{\mu}$, i.e., at any given redshift $T_d \propto M_d^{1/(4+\beta-\gamma)}$, that is $T_d \propto M_d^{-0.38\pm0.04}$, which is expected to be free of selection bias. The dust heating is predominantly fueled by the luminosity of massive ($> 8\,M_\odot$) stars, and so we find that the number of high-mass stars $N_{\rm HM} \propto T_d^{\gamma}$, assuming that the high-mass IMF and stellar content of all SMGs is similar. We may estimate the high-mass star formation efficiency in SMGs by $\eta_{\rm HM} \propto N_{\rm HM}/M_d \propto M_d^{-\gamma / (4+\beta-\gamma)}$; i.e., $\eta_{\rm HM} \propto M_d^{-1.10 \pm 0.13}$. 

Surprisingly, this implies that more dusty galaxies produce, and thus contain, fewer massive stars relative to dust mass. Perhaps this is due to more dusty environments favoring lower mass star formation. This conclusion is consistent with the evolution of $q$ from lower values in SMGs to higher values in nearby dusty disk galaxies \citep{Yun2001} and with the dominance of high-mass stars in SMGs giving way to more radio-quiet low-mass stars heating dust in the local universe \citep{Blain1999b, Baugh2005}.

\section{Scaling Relations}

\begin{figure}
\centering
\includegraphics[height=0.29\textheight]{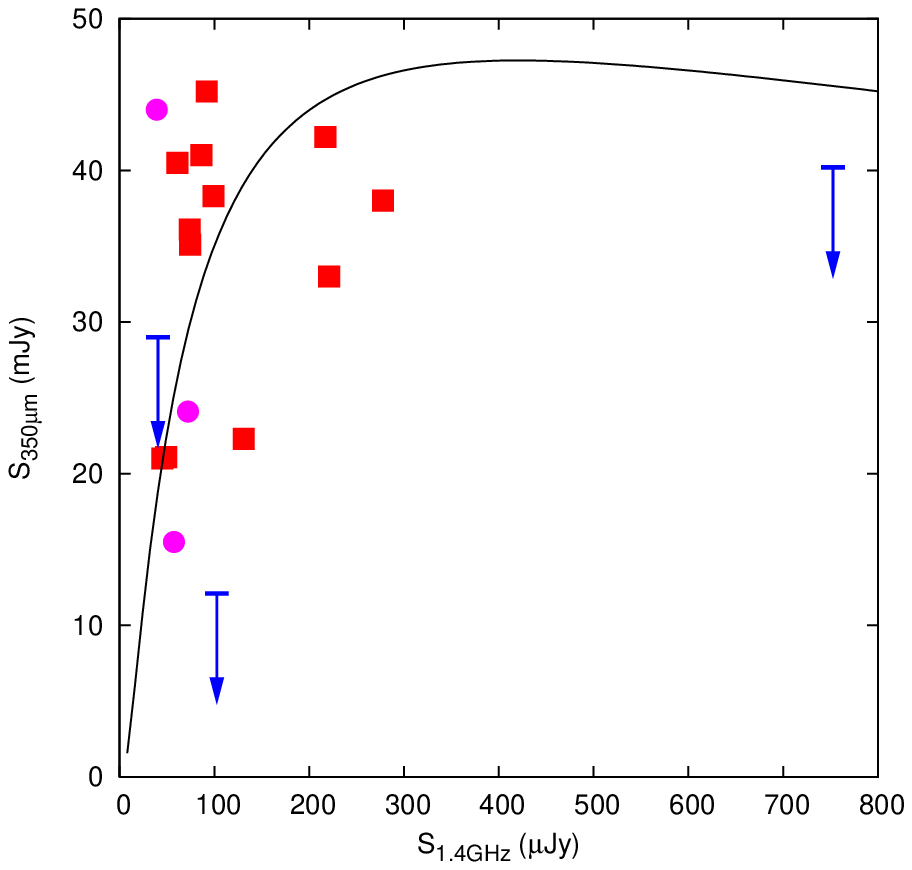}
\includegraphics[height=0.29\textheight]{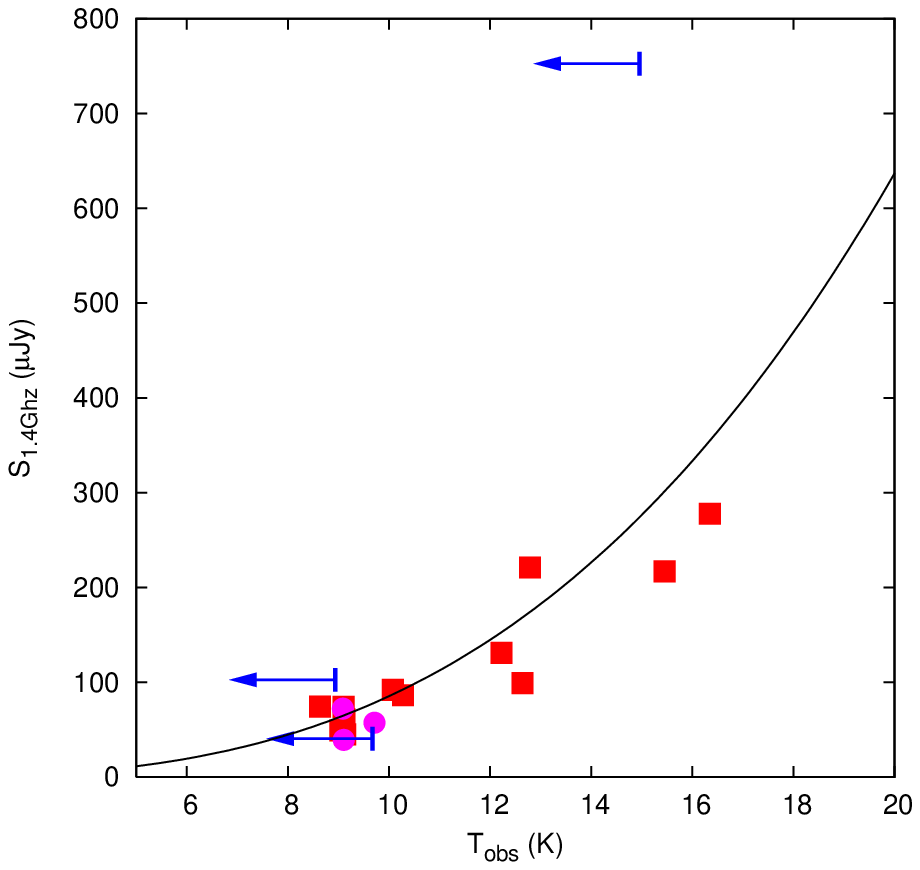}
\includegraphics[height=0.29\textheight]{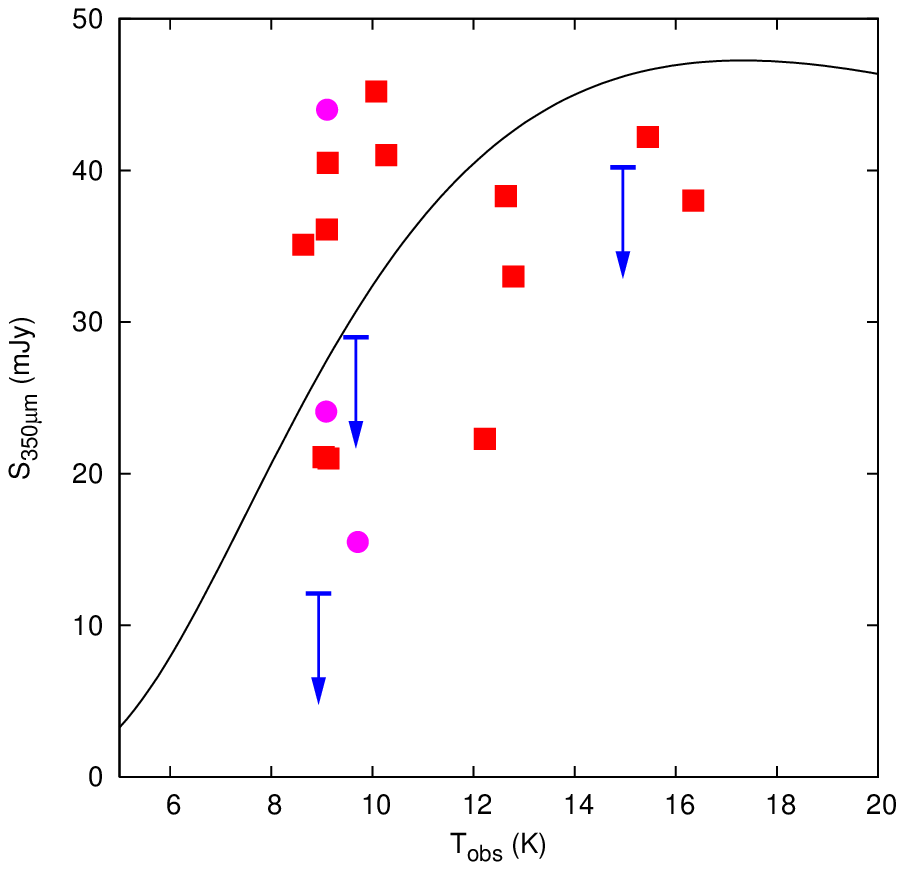}
\caption{
Relations between observed quantities. ({\it a}) Radio and far-infrared fluxes, ({\it b}) radio fluxes, and ({\it c}) far-infrared fluxes vs.\ observed temperatures. Results from this paper are plotted as squares with upper limits indicated as appropriate ({\it arrows}), \citet{Laurent2006} data are shown as circles. The model deriving from the radio-FIR correlation and the observed $L$-$T$ relation is also shown ({\it solid curve}). The data show excellent agreement with predictions within the measurement uncertainties involved. A single radio or submillimeter measurement may, therefore, suffice to predict the other quantities. Thus, these relationship can offer a useful tool for all radio or submillimeter surveys.
}
\label{fig:observed-relations}
\end{figure}

After cancelling the nearly identical redshift terms appearing in the luminosity scaling relations (eq.~[\ref{eq:z-dependance}]), we obtain a useful set of expressions interrelating the observed quantities, $S_{1.4\,{\rm GHz}}$, $S_{\nu}$ and $T_{\rm obs}$ in the far-infrared. For distant SMGs, in the redshift range $z\sim0.5$--4, we find
\begin{eqnarray*}
S_{\rm 1.4\,GHz} & \propto & T_{\rm obs}^{\gamma}, \\
S_{\nu} & \propto & T_{\rm obs}^{\gamma - (4+\beta)} \left( e^{h\nu / kT} - 1 \right).
\end{eqnarray*}
The appropriate constants of proportionality are easily derived from the relevant expressions (eqs.~[\ref{eq:Lfir}], [\ref{eq:Lrad}], [\ref{eq:qL}] and [\ref{eq:L-T}]) applied at the median redshift of the sample. Figure~\ref{fig:observed-relations} testifies to the predictive power of these relations. Excluding the radio-loud points, which clearly lie away from the far-infrared to radio correlation, the model is consistent with observations within the measurement uncertainties. We emphasize, that these relations are not fitted to the data, but rather a direct consequence of the radio to far-infrared correlation, and the deduced $L$-$T$ relation.\footnote{A pure $L$-$T$ relation without the evolution term yields a poor model of the observed relations which therefore directly support the inclusion of the explicit redshift evolution in eq.~(\ref{eq:L-T}).} Therefore, the excellent agreement with data is remarkable.

\section{Conclusions}

New 350\,$\mu$m data lead to the first direct measures of the characteristic dust temperatures and far-infrared luminosities of SMGs and the following valuable conclusions.

\begin{enumerate}

\item The linear radio to far-infared correlation remains valid out to redshifts of $z \sim 1$--3 for SMGs, with the exception of rare radio-loud objects. The power-law index in the correlation is $1.02\pm0.12$, and tighter than observed locally, with an intrinsic dispersion of only $0.12$\,dex. 

\item Either the far--infrared-radio correlation constant $q$ is lower than locally measured ($q_L \approx 2.14\pm0.07$) or the effective dust emissivity index is less, with $\beta \rightarrow 1$ consistent with observations. The lower $q$-value indicates that dust heating is dominated by high-mass stars and that any AGN contribution is minor in comparison.

\item Compared with low-redshift galaxies, SMGs are characterized either by low gas-to-dust ratios, around $54^{+14}_{- 11} (\kappa_{850\,\mu{\rm m}} / 0.15\,$m$^2\,$kg$^{- 1})$, indicating dust-rich environments, or by efficient dust absorption of $\kappa_{850\,\mu{\rm m}} \gtrsim 0.33$\,m$^2\,$kg$^{- 1}$. 

\item Far-infrared- and radio-based photometric redshifts might be appropriate for up to 80\% of SMGs in the redshift range of $z\sim 1.5$--3.5, with a rest-frame temperature assumption of $34.6\pm3.0\,$K$( 1.5/\beta )^{0.71}$. However, photometric redshift indicators may not be appropriate for an unbiased SMG sample, as existing selection effects tend to favor measuring SEDs that are similar in the observed frame.

\item We deduce an $L$-$T$ relationship that we argue to be a property of star-forming galaxies. When recast as mass scaling, we find, that ISM heating scales inversely with the quantity of dust in the galaxies i.e., the galaxies with more dust are less active, possibly due to the formation of more low-mass stars being favoured in dustier environments.

\item From the observed $L$-$T$ relation, possibly biased by selection, and the radio to far-infrared correlation, we derive scaling relations among the observed quantities $S_{1.4\,\rm{GHz}}$, $S_\nu$ in the submillimeter or far-infrared, and the observing frame dust temperature $T_{\rm obs}$, applicable to the redshift range of $z\sim 0.5$--4. A determination of one of these quantities may sufficiently characterise the observed far-infrared properties of SMGs.

\end{enumerate}

\acknowledgements

The authors wish to thank Sophia Khan, Rick Shafer, and Min Yang for their help with collecting data debugging CRUSH; Jonathan Bird for his contributions to the calibration and observing; Thomas Greve and Harvey Moseley for discussions and helpful insights regarding interpretation; Melanie Leong for her efforts to shorten observing times through improvements to the dish surface, Hiroshige Yoshida for all-around software support at the CSO; and the referee for improving this paper through thoughtful comments. We would further like to express our gratitude for the generous sponsorship of the National Science Foundation in funding this research.

{\it Facilities:} \facility{CSO (SHARC-2)}

\clearpage

\end{document}